\documentstyle[12pt,epsfig]{article}
\begin{document}

\date{November 7, 1995}

\title{
\rightline{\small UL--NTZ 29/95}
The equivalent photon approximation for coherent processes at colliders}

\author{R.~Engel$^1$, A.~Schiller$^1$ and  V.G.~Serbo$^{1,2}$  \\
{\it $^1$ Institut f\"ur Theoretische Physik, Universit\"at
Leipzig,}\\{\it D-04109 Leipzig, F.R. Germany}  \\
{\it $^2$ Novosibirsk State University, 630090 Novosibirsk, Russia}}

\maketitle

\begin{abstract}
We consider coherent electromagnetic processes for colliders
with short bunches, in particular the coherent bremsstrahlung (CBS).
CBS is the radiation of one bunch particles  in the collective field
of the oncoming bunch.
It can be a potential tool for optimizing collisions and for measuring
beam parameters. A new simple and transparent method to calculate CBS
is presented based on the equivalent photon approximation for this
collective field.  The results are applied  to the $\phi$--factory
$DA\Phi NE$. For this collider about $ 5 \cdot 10^{14} d E_\gamma / E_\gamma$
photons per second are expected in the photon energy $E_\gamma$ range
from the visible light up to 25 eV.
\end{abstract}

\section{Introduction}

One of the important processes at colliding beams is the
bremsstrahlung. In the last years, besides of the
well-known ordinary (incoherent) bremsstrahlung the coherent
radiation at colliders
has been widely discussed \cite{Bassetti}--\cite{Serbo}.

For definiteness, we consider the  photon emission by electrons
moving through a proton bunch. The ordinary bremsstrahlung dominates
for large enough photon energies.
If the photon energy becomes sufficiently small, the radiation is
determined by the interaction of the electrons with the collective
electromagnetic field of the proton bunch.
It is known (see,
e.g. \S 77 in \cite{Landau2}) that the properties of this coherent
radiation are quite different for electron
deflection angles $\theta_d$ much larger or much smaller than the
typical radiation angle $\theta _r \sim
1/\gamma_e$ where $\gamma_e$ is the electron Lorentz factor.

It is easy to estimate the ratio of these angles\footnote{Throughout
 the paper we use the following notations: $N_e$ and
$N_p$ are the numbers of electrons and protons in the bunches,
$\sigma_z=l$ is the longitudinal, $\sigma_x$ and
$\sigma_y$ are the horizontal and vertical transverse sizes of
the proton bunch, $\gamma_e=E_e/(m_ec^2)$ and
$r_e=e^2/(m_e c^2)$ is the classical electron radius.}.
The electric {\bf E} and magnetic {\bf B} fields of the proton
bunch are approximately equal in magnitude, $|{\bf E}|
\approx  |{\bf B}| \sim eN_p /(l \sigma_x)$.
These fields are transverse and
they deflect the electron into the same direction. In such fields the
electron moves around a circumference of  radius $R\sim \gamma_e
m_e c^2/(eB)$ and gets the deflection angle $\theta_d \sim l/R$.
On the other hand, the  radiation angle $\theta_r$ corresponds to a
length $l_R=R/\gamma_e \sim m_e c^2/(eB)$.
Therefore, the ratio of these angles is determined by the parameter $\eta$
\begin{equation}
\eta={r_e N_p\over \sigma_x} \sim {\theta_d \over \theta _r }
\sim {l \over l_R} \ .
\label{1}
\end{equation}

Let us call the  proton bunch {\it long} if  $\eta \gg 1$.  The corresponding
radiation  is usually called {\it beamstrahlung}.
Its properties are similar to those of the ordinary synchrotron
radiation in an uniform magnetic field (see, e.g.  \cite{Chen}).

The proton bunch is called {\it short} if $\eta \ll 1$. In this
case  the electron trajectory remains practically unchanged
during  the collision.
 In some respect, the radiation in the short bunch fields
 is similar to the ordinary
bremsstrahlung, therefore we call it {\it coherent
bremsstrahlung } (CBS). It differs substantially from the
beamstrahlung.
In  most of the colliders the ratio $\eta$ is either much smaller
than one (all the $pp$,
$\bar p p$ and relativistic heavy-ion colliders, some  $e^+e^-$
colliders and   B--factories) or of the order of one (e.g.  LEP,
TRISTAN). Only  linear $e^+e^-$
colliders have $\eta \gg 1$. Therefore, the CBS has a very
wide region of applicability.

In the following we discuss for our example the characteristic
features of the CBS.
In the usual brems\-strahlung the number of photons emitted by
electrons is proportional to the number of electrons and
protons:
\begin{equation}
dN_{\gamma}\; \propto \;N_e\; N_p\; {dE_{\gamma} \over E_{\gamma}} \ .
\label{2}
\end{equation}
With decreasing photon energies the coherence length
$\sim 4\gamma ^2_e \hbar c /E_\gamma $ becomes comparable to
the length of the proton bunch $l$.  At photon energies
\begin{equation}
E_{\gamma} \stackrel{<}{\sim} E_c= 4 {\gamma^2_e \hbar c \over l}
\label{233}
\end{equation}
 the radiation
 arises from the
interaction of the electron with the proton bunch as a whole,
but not with each proton separately.
The quantity $E_c$ is called the critical photon energy.
Therefore the
proton bunch is similar to a ``particle'' with  the huge charge
$e  N_p$ and with an internal structure described by the form
factor of the bunch. The radiation probability is
proportional to $N_p^2$ and the number of the emitted photons is
given by
\begin{equation}
dN_{\gamma}\; \propto \;N_e\; N^2_p\; {dE_{\gamma} \over E_ {\gamma}} \ .
\label{3}
\end{equation}
The CBS differs strongly from the beamstrahlung in the  soft
part of its spectrum. As one can see from (\ref{3})
 the total number of CBS photons diverges
 in contrast to the beamstrahlung for which (as well as
for the synchrotron radiation) the total number of photons is
finite.

It is useful to discuss shortly the experimental status and
possible applications of this new kind of radiation.
The ordinary bremsstrahlung is a well-known process. Its large
cross section and small angular spread of photons allows  to
use this radiation for measuring one of the important parameters
of a collider --  the luminosity (for example, at  HERA and LEP).
The beamstrahlung has been observed in a single experiment at SLC
\cite{Bon}, and it has been demonstrated that it can be used for
measuring the transverse bunch size.
The main characteristics of the CBS have been calculated
recently \cite{Ginz}--\cite{Serbo}, and
an experiment  for its observation is planned in Novosibirsk.
It seems that CBS can be a potential tool  for optimizing collisions and for
measuring beam parameters. Indeed, the bunch length $l$ can be found from
the CBS spectrum, because the critical energy (\ref{233}) is proportional to
$1 / l$; the horizontal transverse bunch size $\sigma_x$ is related to
the photon rate
$d N_\gamma \propto 1 /\sigma_x^2$.
Besides, CBS may be very useful for a fast control
over an impact parameter between the colliding bunch axes because a dependence
 of the photon rate on this parameter has a very specific behaviour.

As examples we give here the numbers of CBS photons for a
single bunch collision at the $pp$ collider LHC
\begin{equation}
dN_{\gamma }\sim 10^4 {dE_{\gamma}\over
E_{\gamma}} \ \ \ \mbox{at} \ \ \ E_\gamma \stackrel{<}{\sim}
E_c = 590 \  \mbox{eV} \
\end{equation}
and at the B--factory KEKB (for $E_e = 3.5 $ GeV)
\begin{equation}
dN_{\gamma }\sim 10^7 {dE_{\gamma}\over
E_{\gamma}} \ \ \ \mbox{at} \ \ \ E_\gamma
\stackrel{<}{\sim} E_c = 7 \ \mbox{keV} \ .
\end{equation}

A classical approach to CBS was given in \cite{Bassetti}. A quantum
treatment of CBS based on the rigorous concept of colliding wave
packets and some applications of CBS to modern
colliders were  considered in \cite{Ginz}--\cite{Serbo}.
 In this
paper we present a new method to calculate the coherent bremsstrahlung
at colliding beams based on the equivalent
photon approximation (EPA) for the collective electromagnetic
field of the oncoming bunch. This method is much more simple and
transparent as that previously discussed. Our method  allows to
calculate not only the classical radiation but  to take into account
quantum effects in CBS as well. Here  we restrict ourselves to
applications in the classical
limit. CBS is also interesting for relativistic heavy ion colliders,
this topic will be discussed in a forthcoming
publication \cite{Leipzig}.
 Quantum effects for CBS and coherent pair production will be
considered elsewhere \cite{Leipzig2}.

In section 2 we present a qualitative description of the effect.
In the next section
we derive the expressions for the collective field of a charged bunch
as well as the number of equivalent photons.
The luminosity and the polarization are discussed in section 4
followed by expressions for the
spectrum in the next chapter.
In section 6 we apply the obtained formulae for CBS to the case
of the $\phi$--factory
$D A \Phi N E$.

\section{ Qualitative description of CBS }

We start with the standard calculation of the bremsstrahlung (see
\cite{BLP}, \S 93 and \S 97) at
 $ep$ collisions. This process is
defined by the block diagram of Fig. 1a, which gives the
radiation of the electron.
\begin{figure}[htb]
\hspace*{-0.3cm}\epsfig{file=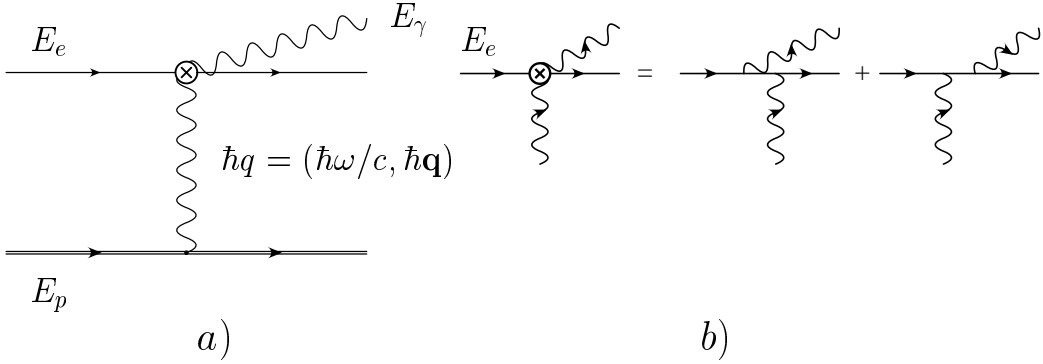,width=15cm}
\caption{Feynman diagrams for the $ep\to ep\gamma$ process: a)
block diagram for bremsstrahlung photon emission by the electron,
b) Compton scattering of equivalent photons on the electron.
\label{cbs1} }
\end{figure}
We denote by $\hbar q$ the 4--momentum of the virtual photon.
 The main contribution to the cross section is given
 by the region of small   values $(- q^2)$.  In this region the
 given reaction can be represented as a Compton scattering of the
equivalent photon  (radiated by the proton) on the electron - Fig.
1b. Therefore, one obtains
\begin{equation}
d\sigma_{ep\to ep\gamma}= dN_{EP} (\omega, E_p)\, d\sigma_{e \gamma}
(\omega,E_e, E_\gamma).
\label{4.2}
\end{equation}
Here $dN_{EP}$ is the number of equivalent photons (EP)
generated by the proton
\begin{eqnarray}
dN_{EP} (\omega, E_p) &\approx &{\alpha \over \pi} {d\omega\over
\omega} \int^{(-q^2)_{\max}} _{(-q^2)_{\min}} {d(-q^2)\over (-q^2)}
\nonumber\\
&\approx&
{\alpha \over \pi}{d \omega\over \omega} \ln{m^2_e\over(m_p\hbar\omega
/E_p)^2}\ ,
\label{4.3}
\end{eqnarray}
where
\begin{equation}
(-q^2)_{\min}={(m_p c)^2 \omega^2   \over E_p(E_p- \hbar \omega)}\approx
\left( {m_p c   \omega\over E_p} \right)^2
\end{equation}
is determined by kinematics and
$(-q^2)_{\max}\sim (m_e c/\hbar)^2$ is given by the dynamical cut--off
of the Compton cross section.
Therefore, the accuracy of this approximation is logarithmic.

The Compton cross section is of the order of
\begin{equation}
d\sigma_{e\gamma} \sim 2\pi \alpha r^2_e\; {\omega_{\min} \over \omega}
{dE_\gamma\over E_\gamma}
\label{4.7}
\end{equation}
with the minimal energy of the equivalent photon
\begin{equation}
\hbar \omega_{\min}={m^2_e c^4\over 4E_e(E_e-E_\gamma)}\;E_\gamma \ .
\label{4.4}
\end{equation}
For the cross section (\ref{4.2}) we obtain the estimate
\begin{eqnarray}
d\sigma_{ep\to ep\gamma}&\sim &4\alpha r^2_e {dE_\gamma\over
E_\gamma} \int^\infty_{\omega_{\min}} \; d\omega\; {\omega_{\min}
\over\omega^2}\ln{E_p m_e\over m_p \hbar \omega}
\nonumber\\
&\sim& 4\alpha r^2_e
\ln{{4E_e E_p (E_e-E_\gamma)\over m_e m_p
c^4\,E_\gamma}} \; {dE_\gamma\over E_\gamma}\ .
\label{4.8}
\end{eqnarray}

Just as in the standard calculations we can estimate the number of CBS photons
using EPA.  Taking into
account that the number of EP increases by a factor $\sim N_p$
compared to the ordinary bremsstrahlung we get (using $ d(-q^2) \to d^2
q_\bot / \pi$)
\begin{equation}
dN_{EP} \sim N_p \; {\alpha \over \pi^2}\; {d\omega \over
\omega} \; {d^2 q_\bot \over q_\bot^2}.
\label{4.24a}
\end{equation}

 Since the impact parameter {\mbox{\boldmath$\varrho$}}
is of the order of
\begin{equation}
\varrho_x \sim {1 \over q_x}, \ \
\varrho_y \sim {1 \over q_y}
\label{4.10}
\end{equation}
we can rewrite this expression in another
form
\begin{equation}
dN_{EP} \sim N_p \; {\alpha \over \pi^2}\; {d\omega \over
\omega} \; {d^2 \varrho \over \varrho^2} \ .
\label{4.24b}
\end{equation}

To perform the integration over {\mbox{\boldmath$\varrho$}} we
 estimate the main
integration region as follows.
The region of small impact paramaters
 $|\varrho_x| \ll
\sigma_x$ and $|\varrho_y| \ll \sigma_y$
does not give a significant contribution because
the electromagnetic field vanishes in the centre of a symmetric bunch.  The
region of large impact parameter
$|\varrho_x| \gg \sigma_x$ and $|\varrho_y| \gg \sigma_y$
also does not contribute significantly
since the proton bunch form factor decreases rapidly.
 Therefore, the
region which gives the main contribution  is
\begin{equation}
|\varrho_x | \sim  |1/q_x| \sim \sigma_x, \;\;\;
|\varrho_y| \sim |1/q_y| \sim  \sigma_y \ .
\label{4.25}
\end{equation}
Integrating over this region we obtain
\begin{equation}
dN_{EP} \sim N_p \; {\alpha \over \pi}\; {d\omega \over
\omega} \;{\sigma_x \sigma _y \over \sigma_x^2 +\sigma_y^2 } \ .
\label{4.26}
\end{equation}
As a result
 the ``effective cross section"$\;$  for CBS is of the order of
\begin{equation}
d \sigma _{\rm eff}  \sim N_p\, \alpha\, r_e^2\; {\sigma_x \sigma _y
\over \sigma_x^2 +\sigma_y^2 } {dE_\gamma \over E_\gamma}.
\label{4.27}
\end{equation}
In particular, for flat beams $\sigma_y \ll \sigma_x$
one has
\begin{equation}
d \sigma _{\rm eff}  \sim N_p\, \alpha\, r_e^2\; {\sigma_y
\over \sigma_x } {dE_\gamma \over E_\gamma}.
\label{271}
\end{equation}

To illustrate the transition from the ordinary bremsstrahlung to CBS
we present in Fig. 2
qualitatively the photon spectrum for the HERA collider
using the results of the last paper in \cite{Ginz}.
\begin{figure}[htb]
\epsfig{file=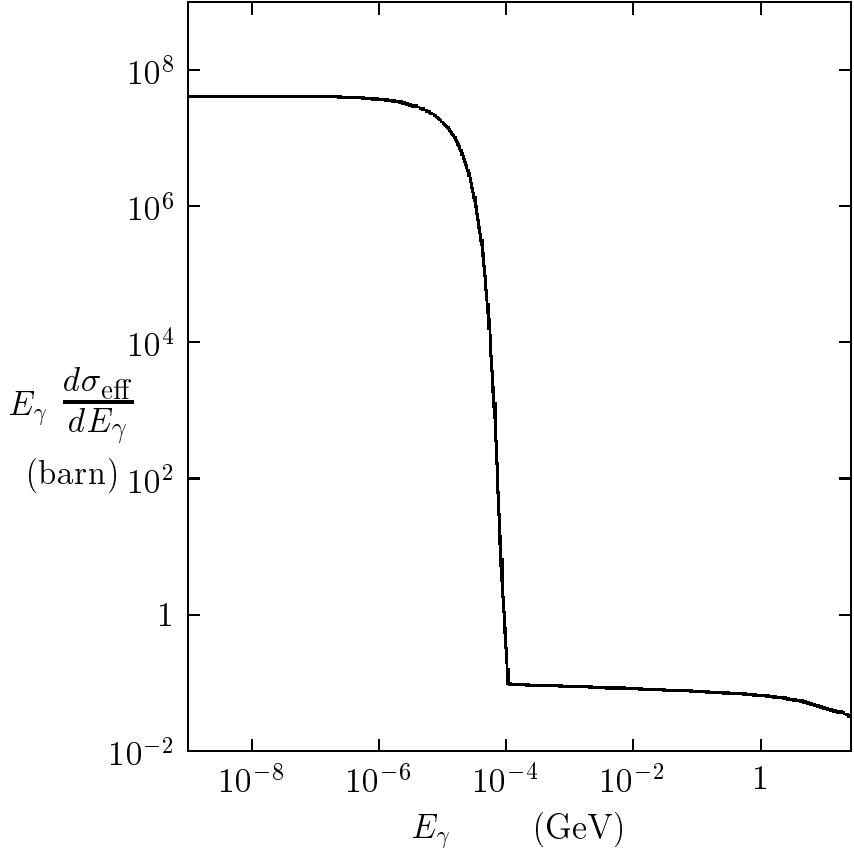,width=12cm}
\caption{Effective cross section for the emission of
bremsstrahlung photons at   HERA as function of the photon energy.
The huge increase at low
photon energies is due to the coherent bremsstrahlung effect, the high energy
tail corresponds to ordinary (incoherent) bremsstrahlung.
\label{cbs2} }
\end{figure}
In the region of $E_\gamma \sim 10$ keV
the number of photons dramatically increases by about 9 orders of magnitude.

\section{The collective electromagnetic field of a charged bunch and
the number of equivalent photons}

To define a coordinate frame
we choose the $z$--axis along the momentum of the initial electron,
the $x$-- and $y$--axes in the transverse horizontal and
vertical directions, respectively.
The possible changes
in the transverse sizes of the bunches during
the collision are neglected.

The electromagnetic field of a single  proton moving with the
constant velocity ${\bf v}$ (along $(-z)$--axis)
can be found for example in \cite{Landau2}, \S 64.
An expansion of this field in
plane waves has the form
\begin{eqnarray}
{\bf E_q}(t)&=&-4\pi i  e {{\bf q}-{\bf v}({\bf qv})/c^2 \over {\bf
q}^2 - ({\bf qv}/c)^2} \ \mbox{e}^{-i{\bf qv}t};
\nonumber\\
{\bf B_q}(t)&=&{{\bf v}\over c }\times {\bf E_q}(t).
\label{7}
\end{eqnarray}
{}From that expression one can see that the wave with
the wave vector  ${\bf q}$ has the frequency
\begin{equation}
\omega = {\bf qv} \ .
\label{8}
\end{equation}
For such a wave we put in correspondence an equivalent photon   with the
4-momentum $\hbar q = (\hbar \omega/c,\hbar {\bf q})$.

Now we consider the field of the whole bunch of protons. If the particle
distribution  does not change considerably during
the collision, the proton density
$n_p ({\bf r},t)$ depends on  ${\bf r}$ and $t$ in the
following combination  only
\begin{eqnarray}
n_p ({\bf r},t)&=&n_p ({\bf r}-{\bf
v}t)= n_p({\mbox{\boldmath $\varrho$}}, z+vt);
\nonumber\\
{\bf r}&=&(
\mbox {\boldmath$\varrho$},z)=(\varrho_x,\varrho_y,z) \ .
\label{9}
\end{eqnarray}

Let us define the form factor of the proton bunch density at $ t=0$
\begin{equation}
F_p({\bf q})= \int n_p({\bf r})\;\mbox {e}^{-i{\bf qr}}\; d^3r
\end{equation}
with its normalization
\begin{equation}
F_p(0)=\int n_p({\bf r})\; d^3r =N_p \ .
\end{equation}
We find the electric and
 magnetic fields of the proton bunch analogous to the calculation in
the problem of \S 64 in \cite{Landau2}
\begin{eqnarray}
{\bf E}({\bf r},t)&=&\int {\bf E_q}(t) \;F_p({\bf q}) \ \mbox
{e}^{i{\bf qr}}\; {d^3 q \over (2\pi )^3};\nonumber\\
{\bf B}({\bf
r},t)&=&{{\bf v}\over c}\times {\bf E}({\bf r},t)
\label{10}
\end{eqnarray}
where  ${\bf E_q}(t)$ is the same as in (\ref{7}).
The spectral expansion of the proton bunch field contains
the same frequencies (\ref{8}). The spectral components of these fields are
\begin{equation}
{\bf B}_\omega ({\bf r})= \int {\bf B}({\bf r},t) \
\mbox{e}^{i\omega t}\;dt = \frac{{\bf v}}{c}\times {\bf E}_\omega ({\bf
r})\ ;
\end{equation}
\begin{eqnarray}
{\bf E}_\omega ({\bf r})&=&- {4\pi i e \over v} \ \mbox
{e}^{iq_z z} \nonumber\\
&\times& \int {{\bf q}-{\bf v}\omega /c^2 \over {\bf
q}_\bot^2 + \omega^2 / (v\gamma _p)^2}\;F_p({\bf q}) \ \mbox {e}^
{i{\bf q}_\bot \mbox{\scriptsize\boldmath $\varrho$}}\;
{d^2 q_\bot \over (2\pi)^2};
\nonumber\\
\gamma _p &=&{1\over \sqrt {1-(v/c)^2}}.
\label{11}
\end{eqnarray}

The form factor $F_p({\bf q})$ decreases rapidly when the
components of ${\bf q}$ become larger than the inverse
proton bunch sizes. This means that the characteristic values of
$\omega /(v\gamma _p) \sim 1/ (l\gamma_p)$ are much less than those
 of  $q_\bot \sim 1/ \sigma_y$. Therefore,
one can omit the quantity $\omega^2 /(v\gamma _p)^2$ in the integrand of
(\ref{11}).

The electromagnetic field is dominated by its transverse components.
For the transverse electric field
\begin{eqnarray}
{\bf E}_\bot ({\bf r},t)=-4\pi i e \int {{\bf q}_\bot
\over {\bf q}^2 - (\omega /c)^2}\;F_p({\bf q}) \
\mbox {e}^{i({\bf q r}- \omega t)}\; {d^3 q \over (2\pi )^3}
\nonumber\\
\label{12a}
\end{eqnarray}
its spectral component (in which we omit an irrelevant phase
factor exp$(iq_z z)$ ) has the form
\begin{equation}
{\bf E}_\omega ({\mbox{\boldmath $\varrho$}})=- {4\pi i e \over v} \ \int
{{\bf q}_\bot \over {\bf q}^2_\bot }\;F_p({\bf q}) \ \mbox
{e}^{i{\bf q}_\bot {\mbox{\scriptsize\boldmath $\varrho$}}}\;
{d^2 q_\bot \over (2\pi)^2} \ .
\label{12b}
\end{equation}
The field ${\bf E}_\bot ({\bf r},t)$ is real, hence    $|{\bf
E}_{-\omega} ({\mbox{\boldmath $\varrho$}})|=
|{\bf E}_{\omega}^* ({\mbox{\boldmath $\varrho$}})|$
from which we get the  relation
\begin{equation}
\int ^\infty _{-\infty} \mid {\bf E}_\bot
({\bf r},t) \mid ^2\; dt = 2\int ^\infty _0 \mid {\bf E}_\omega (
\mbox{\boldmath$\varrho$})\mid ^2\;{d\omega \over 2\pi}.
\label{13}
\end{equation}

Having calculated the electromagnetic field of the proton bunch we
calculate now the number of equivalent photons.
The main idea of EPA is that the
electromagnetic interaction of an electron with the complicated
field of the proton bunch is replaced by a more simple Compton scattering
of this electron with the flux of EP generated by the proton bunch.

 To perform such a
reduction, let us remind a few facts about the collisions of two
beams.
It is well--known (see,  e.g. \S 12 in \cite{Landau2}) that a number of
events for a process with the cross section $\sigma _{12}$ is
equal to
\begin{equation}
dN =dL_{12} \ d\sigma _{12};  \;\;dL_{12}=(v_1+v_2)n_1
n_2\;d^3r\;dt
\label{14}
\end{equation}
where $n_j$ is the particle density of the $j$-th bunch, $v_j$
 its velocity and $L_{12}$  the luminosity for a single
head--on collision of the bunches. If these densities are of the
form  (\ref{9})
\begin{equation}
n_1=n_1 ({\mbox{\boldmath $\varrho$}}, z-v_1t);\;\;
n_2=n_2 ({\mbox{\boldmath $\varrho$}},
z+v_2t)
\label{15}
\end{equation}
one can  replace  $z,\;t$ by the new variables $z-v_1t,\;
z+v_2t$ and  integrate the expression (\ref{14})  independently
 over these new variables. After these integrations the luminosity depends
on the
so called ``transverse densities''
\begin{equation}
n_j ({\mbox{\boldmath $\varrho$}})= \int n_j\;dz \ .
\label{16}
\end{equation}
The transverse density is equal to the total number of corresponding
particles which cross a unit
area around the impact parameter
${\mbox{\boldmath $\varrho$}}$ during the collision.
As a result, the luminosity is
\begin{equation}
dL_{12}=n_1 ({\mbox{\boldmath $\varrho$}})
\;n_2 ({\mbox{\boldmath $\varrho$}})\;d^2\varrho \ .
\label{17}
\end{equation}

Now we can apply these formulae for CBS  considering it
as the scattering of
electrons (index 1) on the  electromagnetic field of the proton
bunch. Replacing this field by the flux of EP  (index 2) with
some spectrum we obtain
the number of the produced CBS photons in the form
\begin{eqnarray}
dN_\gamma &=& dL_{e \gamma }(\omega ) \  d\sigma_{e \gamma}
(\omega ,E_\gamma );\nonumber\\
dL_{e \gamma }(\omega )&=& n_e ({\mbox{\boldmath $\varrho$}}) \ n_\gamma
({\mbox{\boldmath $\varrho$}}, \omega) \ d\omega \  \; d^2\varrho \ .
\label{18}
\end{eqnarray}
Here $n_e ({\mbox{\boldmath $\varrho$}})$ is the transverse electron density,
$n_\gamma ({\mbox{\boldmath $\varrho$}},\omega) d\omega $  the transverse
density of EP with frequencies in the interval from $\omega$
to $\omega + d\omega $. $dL_{e \gamma }(\omega )$ denotes the differential
luminosity for the collisions of  electrons  and EP and $d \sigma _
{e \gamma } (\omega, E_\gamma)$ is the cross section of the
Compton scattering of the electron on the equivalent photon with the frequency
$\omega $.

In our case of ultrarelativistic
 protons ($\gamma_p \gg 1$, $v \approx c$) the electric and magnetic fields
of the proton bunch (\ref{10}) are approximately equal, transverse
and orthogonal to each other
\begin{equation}
{\bf E} \approx {\bf E}_\bot \ , \ \  {\bf B} \approx {\bf B}_\bot \ , \ \
|{\bf E}| \approx  |{\bf B}| \ , \ \  {\bf E \cdot B} = 0 \ .
\label{19}
\end{equation}
Therefore, they are similar to the usual fields describing  waves of light.
We obtain for the flux of the electromagnetic energy through the
unit area around the impact
parameter ${\mbox{\boldmath $\varrho$}}$ the known expression
\begin{equation}
{c\over 4\pi}\; \int ^\infty _{-\infty} \mid {\bf E}_\bot ({\bf
r},t) \mid ^2\; dt \ .
\end{equation}
It can be rewritten as the total energy of EP
crossing the same area during the collision
\begin{equation}
\int ^\infty _0 \hbar \omega \ n_\gamma
({\mbox{\boldmath $\varrho$}}, \omega)\;d\omega \ .
\end{equation}
Using (\ref{13})
we obtain the transverse density of the EP
\begin{eqnarray}
n_\gamma ({\mbox{\boldmath$\varrho$}},
 \omega )\;d\omega &=&{c\over 4\pi^2}\; \mid {\bf E}_\omega
({\mbox{\boldmath $\varrho$}})\mid ^2\;{d\omega \over \hbar \omega}
\nonumber\\
&=& 4\alpha\;{d\omega\over \omega} \left|
\int { {\bf q}_\bot \over {\bf q}^2_\bot } \;F_p({\bf q}) \
\mbox{e}^{i{\bf q}_\bot
\mbox{\scriptsize\boldmath$\varrho$}}\;
{d^2 q_\bot \over (2\pi)^2}\right|^2.
\label{20}
\end{eqnarray}
It should be noticed that the integration over ${\bf q}_\bot  $
 can be extended to $\infty$ even at $\varrho \to 0$ due to
the proper behaviour of the form factor. It
 provides a high accuracy of our method compared to the usual
 Weizs\"acker--Williams approximation
(discussed in section 2) which is correct only logarithmically.

\section{Luminosity $L_{e \gamma }$ and polarization of EP}

Based on the obtained expression for the equivalent photon number we
can calculate the luminosity $L_{e \gamma }$.
Substituting (\ref{20}) into (\ref{18}) and
performing the integration over $\mbox{\boldmath $\varrho$}$
we  express the luminosity in the form
\begin{equation}
dL_{e \gamma }(\omega) =
{{\alpha}\over{\pi}}{d\omega \over \omega } J(\omega )
\label{210}
\end{equation}
with the function $J(\omega)$ containing both the electron
$F_e$ and the proton $F_p$ bunch form factors
\begin{eqnarray}
J(\omega )=4\pi \int{{\bf q}_{\perp}{\bf q}^{\prime}_{\perp}\over
{\bf q}^{2}_{\perp}{{\bf q}^{\prime}_{\perp}}{}^{2}}\;F_{p}({\bf q})
F^{*}_{p}({\bf q}^\prime)F_{e}({\bf q}^\prime- {\bf q})\;
{d^{2} q_{\perp}d^{2} q^{\prime}_{\perp}\over (2\pi )^{4}}
\nonumber\\
\label{21}
\end{eqnarray}
and
\begin{equation}
q_{z}= q^{\prime}_{z} = - \omega /c.
\label{22}
\end{equation}

It is useful to consider in detail the important case of Gaussian beams
since usually it is assumed that near the interaction point the
particle distribution of the bunches is Gaussian like.
In
this case the density $n_p({\bf r})$ can be represented as a
product of the transverse $n_p({\mbox{\boldmath $\varrho$}})$ and longitudinal
$n_{p }(z)$ densities
\begin{eqnarray}
n_p({\bf r})&=& n_p({\mbox{\boldmath $\varrho$}})\cdot n_{p }(z);\; n_{p}(z)=
{1\over \sqrt{2\pi} \sigma _{pz}}\; \mbox {exp}\left(
-{z^2\over 2\sigma ^2_{pz}}\right);\nonumber\\
l&\equiv& \sigma _{pz};
\end{eqnarray}
\begin{equation}
n_p({\mbox{\boldmath $\varrho$}})= {N_p\over 2\pi \sigma _{px}
\sigma _{py}}\; \mbox {exp} \left( -{\varrho_x^2\over 2\sigma_{px}^2} -
{\varrho_y^2\over 2\sigma_{py}^2} \right) ,
\label{24a}
\end{equation}
and the form factor of the proton bunch is
\begin{eqnarray}
F_p({\bf q})=N_p\; \mbox{exp} \left[ -{1\over 2}(q_x\sigma_{px})^2
-{1\over 2}(q_y\sigma_{py})^2 -
{1\over 2} (\omega l/c)^2 \right].
\nonumber\\
\label{24b}
\end{eqnarray}

Analogous formulae take place for the electron bunch
replacing the index $p\to e$.  In the general case,  when the
 electron bunch axis is
shifted by a distance ${\bf R}=(R_x, R_y)$ from the proton bunch axis
the  electron bunch form factor in Eqs. (\ref{21}), (\ref{23})
 is equal to
\begin{eqnarray}
F_e({\bf q}^\prime -{\bf q})=N_e\; \mbox{exp}
\bigg[ -{1\over 2}(q_x^\prime-q_x)^2 \sigma_{ex}^2
\nonumber\\
-{1\over 2}(q_y^\prime -q_y)^2 \sigma_{ey}^2 -
i ({\bf q}^\prime _\bot - {\bf q}_\bot){\bf R} \bigg] \  .
\label{25}
\end{eqnarray}
 From these
expressions we immediately obtain the important relation
\begin{eqnarray}
{J(\omega)\over J(0)}&=&\left| \int n_{p}(z) \mbox{exp}(-iq_z z)dz
\right| ^2\nonumber\\
&=& \mbox{exp}[-(\omega l/c)^2] \ .
\label{26}
\end{eqnarray}

Now we discuss the polarization of EP.
The local polarization of EP is determined by the
field ${\bf E}_\omega ({\mbox{\boldmath $\varrho$}})$.  In particular, the
unit vector
\begin{equation}
{\bf e} = {{\bf E}_\omega ({\mbox{\boldmath $\varrho$}}) \over
\mid {\bf E}_\omega
({\mbox{\boldmath $\varrho$}})\mid}
\end{equation}
is the local polarization vector of EP, and the matrix $e_i e_k^*$
is the local density matrix of EP.
To obtain the average density matrix one has to integrate this local matrix
with the luminosity (\ref{18}) over the impact parameter
{\mbox{\boldmath $\varrho$}}  and
normalize it. With this procedure we obtain
\begin{equation}
<e_i e_k^*> = {J_{ik} \over J(\omega)};\;\; i,k= x,y;\;\;J(\omega)=
J_{xx} + J_{yy};
\label{266}
\end{equation}
\begin{eqnarray}
J_{ik}=4\pi \int{ q_{i} q^{\prime}_{k}\over {\bf
q}^{2}_{\perp}{{\bf q}^{\prime}_{\perp}}{}^{2}}\;F_{p}({\bf q})
F^{*}_{p}({\bf q}^\prime)F_{e}({\bf q}^\prime- {\bf q})\; {d^{2}
q_{\perp}d^{2} q^{\prime}_{\perp}\over(2\pi )^{4}} \  .
\nonumber\\
\label{23}
\end{eqnarray}
{}From the expression (\ref{266}) we obtain
the average Stokes parameters of the equivalent
photons describing their linear polarization
\begin{equation}
\xi_1= { J_{xy}+ J_{yx} \over J(\omega)} \ ; \ \ \
\xi_3= { J_{xx}- J_{yy} \over J(\omega)} \ .
\label{267}
\end{equation}

Eqs.  (\ref{210}--\ref{22},\ref{266},\ref{23}) are the
basic formulae to calculate  the CBS.
Their accuracy can be estimated in the framework of the
general approach developed in  \cite{Ginzyaf}. Here we only point out
the  necessary conditions  for their application, namely  the
bunches should be short and their sizes should not change
considerably during the collisions.

To obtain the energy and the angular distribution
of the CBS photons and their polarization it is sufficient
to calculate the quantities $J(0)$ and $<e_i e_k^*>$.
Such a calculation for Gaussian beams has been performed
in \cite{Ginzyaf} and below we
will use it.

\section{Spectrum of CBS photons}

To calculate the number  of CBS photons  as given in
Eqs.~(\ref{18},\ref{210}) one has to use
the known Compton cross
section (e.g. , from \cite{Ginznim})
\begin{eqnarray}
d \sigma_{e\gamma} =  2 r_e^2 \ {d E_\gamma \over E_\gamma} \ {
d \varphi \over (1+ z)^3}
\big[ 1 + z^2
\nonumber\\
- 2 z(\xi_3 \cos 2 \varphi - \xi_1 \sin 2 \varphi) \big] \
\label{268}
\end{eqnarray}
where
\begin{equation}
z\;=\; (\gamma_e \theta )^2 \ .
\label{269}
\end{equation}
Here $\theta$ and $\varphi$ are the polar and azimuthal angles of
the CBS photons, respectively.
The EP
energy $\hbar \omega$ is related to the energy $E_\gamma$ and
the emission angle $\theta$ of the CBS photon by a simple kinematical
relation
\begin{equation}
\hbar \omega  = (1+z) {E_{\gamma }\over
4\gamma_{e}^{2}(1-E_{\gamma}/E_{e})}\;  \  .
\label{28}
\end{equation}
Introducing the   constant $N_0$
\begin{equation}
   N_0= {8\over 3}\;\alpha\;r_e^2\;J(0)
\label{30}
\end{equation}
we obtain the angular--energy distribution for the CBS photons
(for  unpolarized electrons and after
integrating over the azimuthal angle $\varphi$)
\begin{eqnarray}
dN_\gamma &=& {3\over 2}N_0 {dE_\gamma \over E_\gamma} {dz\over
(1+z)^2} \nonumber\\
&\times& \left[ {1+z^2\over (1+z)^2} \left( 1-{E_\gamma\over E_e}\right)
+{E_\gamma ^2 \over 2E_e^2 } \right] \  {J(\omega )\over J(0)} \ .
\label{29}
\end{eqnarray}

The energy spectrum of CBS photons is  obtained by integrating (\ref{29})
over $z$
\begin{equation}
dN_\gamma = N_0\;\Phi (E_\gamma,E_e )\; {dE_\gamma \over E_\gamma};
\end{equation}
\begin{eqnarray}
\Phi(E_\gamma,E_e)&=&{3\over 2}
\int_{0}^{\infty}{dz\over (1+z)^2}\nonumber\\
&\times& \left[{1+z^2\over (1+z)^2}\left( 1-{E_\gamma\over E_e}\right)
+{E_\gamma ^2 \over 2E_e^2 } \right]\cdot {J(\omega )\over J(0)} \  .
\label{31a}
\end{eqnarray}
The spectral function  $\Phi(E_\gamma,E_e) $ is normalized by the condition
\begin{equation}
\Phi(0,E_e)=1 \ .
\label{31b}
\end{equation}

The properties of this spectral function are determined by the ratio
$J(\omega )/ J(0)$. For  Gaussian beams this ratio depends on
the longitudinal density of the proton bunch (see Eq. (\ref{26})).
The constant $N_0$ depends  on the transverse densities of
the electron and proton bunches only.

Now we consider the important case of flat Gaussian
beams ($a_y^2 = \sigma_{ey}^2
+\sigma_{py}^2 \ll a_x^2=\sigma_{ex}^2+\sigma_{px}^2$). In that case
the constant $N_0$ is equal to \cite{Ginzyaf}
\begin{eqnarray}
N_0&=&{8\over 3\pi}\;\alpha N_e \left({r_e N_p
\over a_x}\right) ^2\nonumber\\
&\times &{\arcsin (\sigma _{ex}/a_x)^2 + \arcsin
(\sigma _{ey}/a_y)^2 \over [1-(\sigma _{ex}/a_x)^4] ^{1/2}},
\label{77}
\end{eqnarray}
in particular, for  flat and identical beams ($\sigma_{x} = \sigma_{ex,px}$,
$\sigma_{y} = \sigma_{ey,py}$)
\begin{equation}
N_0={8\over 9\sqrt{3}}\;\alpha N_e \left({r_e N_p \over
\sigma _x}\right) ^2.
\label{88}
\end{equation}
Integrating Eq.~(\ref{17}) over $d^2 \varrho$ we find the luminosity of
$ep$ collisions
\begin{eqnarray}
L_{ep}({\bf R})&=&L_{ep}(0) \ \mbox{exp}\left( -{R_x^2\over 2a^2_x}
-{R_y^2\over 2a^2_y} \right);\nonumber\\
L_{ep}(0)&=&{N_eN_p\over 2 \pi a_x a_y}
\label{270}
\end{eqnarray}
and define the ``effective cross section''
\begin{equation}
d \sigma_{\rm eff}= { dN_\gamma \over L_{ep}} \ .
\label{89}
\end{equation}
For the case of identical flat beams and at $R=0$ this quantity is of the form
\begin{equation}
d \sigma _{\rm eff}
= {32 \pi \over 9 \sqrt{3}} N_p\, \alpha\, r_e^2\; {\sigma_y
\over \sigma_x } \ \Phi(E_\gamma,E_e) \ {dE_\gamma \over E_\gamma}
\label{90}
\end{equation}
in accordance with the estimate (\ref{271}).

All obtained formulae are valid both in the classical ($E_c \ll E_e$) and
in the quantum ($E_c \gg E_e$) cases. Below we consider only the classical
limit which is valid for all the existing colliders.
The quantum effects being at the moment of mainly theoretical interest will be
studied in  \cite{Leipzig2}.

In the classical case the    energy of the CBS photons is
$E_\gamma \stackrel{<}{\sim} E_c$.
 The angular--energy distribution (\ref{29})  simplifies to the expression
\begin{equation}
dN_\gamma = {3\over 2}N_0 {dE_\gamma \over E_\gamma}\;
dz\; {1+z^2\over (1+z)^4} \ {J(\omega )\over J(0)};
\end{equation}
\begin{equation}
\omega  = {E_{\gamma }\over 4\gamma_{e}^{2}\hbar}\;
 (1+z) = \frac{E_\gamma}{E_c} \frac{c}{l} (1+z)\ .
\label{36}
\end{equation}

\section{Application of CBS to the $ DA \Phi NE $ collider}

The $DA\Phi NE$ $e^+e^-$
collider has been proposed in 1990 and it is now under construction
at Frascati \cite{FINUDA}.
The collider is planned to
run in 1997 at cms energy of the $\phi$ meson with an initial
luminosity of $10^{32}$  cm$^{-2}$s$^{-1}$.
The expected collider parameters are the following \cite{FINUDA,PDG}
\begin{eqnarray}
N_e &=& N_{e^+} =0.89 \cdot 10^{11} \ , \ \  \sigma_z=3 \ \mbox{cm}\ , \ \
\sigma_x= 2 \ \mbox{mm} \ , \nonumber\\
\sigma_y &=& 0.02 \ \mbox{mm} \ , \ \
E_e=0.5 \ \mbox{GeV} \ .
\label{91}
\end{eqnarray}

The parameter $\eta$ from (\ref{1}) is equal to
\begin{equation}
\eta =0.12
\label{92}
\end{equation}
from which follows that
$DA\Phi NE$ belongs to the colliders with short bunches for which
the photon radiation at small energies is the coherent bremsstrahlung.

\subsection{Spectrum of CBS photons}

Using the formulae (\ref{233}) and (\ref{88}) we find
the critical energy $E_c$ of CBS photons and the constant $N_0$
\begin{equation}
E_c = 25 \ \mbox{eV}  \ , \ \ N_0= 5.2 \cdot 10^6 \ .
\label{921}
\end{equation}

The number of CBS
photons for {\it a single collision} of the beams is
\begin{equation}
dN_{\gamma }=N_{0}\ \Phi (E_{\gamma}/E_{c}){dE_{\gamma}\over
E_{\gamma}}
\label{93}
\end{equation}
where the function $\Phi(x)$ is obtained from (\ref{31a})
in the classical limit $E_c \ll E_e$
\begin{equation}
\Phi (x)={3\over 2}\;\int _0^\infty {1+z^2\over (1+z)^4}\; \mbox
{exp} [-x^2(1+z)^2]\;dz \ ,
\label{94}
\end{equation}

\begin{eqnarray}
\Phi (x)&=&1 \;\;\mbox{at}\;\; x \ll 1;\nonumber\\
\Phi (x)&=&(0.75/x^2)\cdot \mbox{e}^{-x^2} \;\;\mbox{at}\;\; x\gg 1 \ .
\end{eqnarray}
Some values of this function are: $\Phi (x)=$ 0.80, 0.65, 0.36,
0.10, 0.0023 for $x=$ 0.1, 0.2, 0.5, 1, 2.

It may be convenient for $DA\Phi NE$ to use the CBS photons in the
range of the {\it visible light} $E_\gamma \sim 2-3$ eV $\ll$ 25
eV.  In this region the rate of photons is expected to be
\begin{equation}
{dN_\gamma \over \tau} \approx 5 \cdot 10^{14}\; {dE_\gamma \over
E_\gamma} \;\; \;\mbox{photons$\;$ per $\;$ second}
\label{95}
\end{equation}
where $\tau=0.0108 \; \mu$s is the time between collisions of the bunches
at a given interaction region.
Additionally, in this energy region  the photon polarization should be easily
measurable.

To use the CBS photons for monitoring the
beams a special window should be installed
in the beam pipe.
A serious problem for the observation and application of CBS may
be  the background due to synchrotron radiation on the
external magnetic field of the accelerator. This background strongly
depends  on the details of the magnetic layout of the
collider.

\subsection{Collisions with a  non--zero impact parameter of the
bunches}

If the electron bunch  axis  is  shifted  in  the horizontal (vertical)
direction by a distance $R_{x}$ ($R_y$) from the  positron bunch axis,
the luminosity  decreases exponentially (see Eq. (\ref{270})).
On the contrary, in the case of a vertical displacement  the number of
CBS photons increases almost two times (by $ 92 \% $ at $R_{y}=
4.0\,\sigma _{y}$). After that, the rate of photons decreases, but even
at $R_{y}= 15\,\sigma_y$ the normalized photon rate
reaches $\approx 1.65$.
  The corresponding curve is presented in Fig. 3.
\begin{figure}[htb]
\epsfig{file=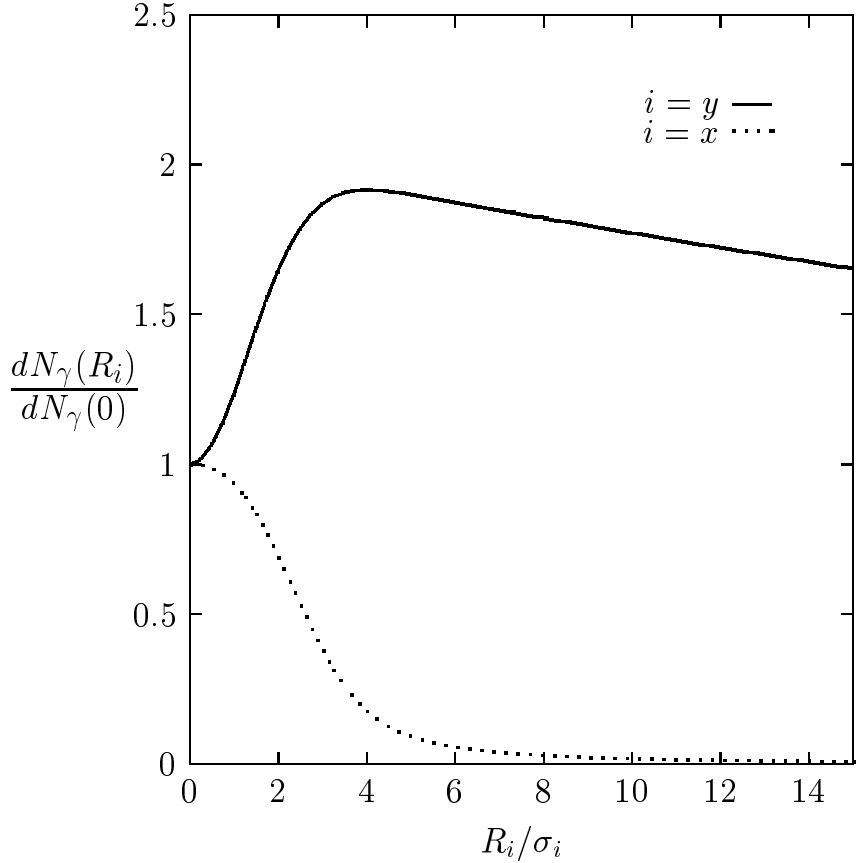,width=12cm}
\caption{Normalized photon emission rate at  $DA\Phi NE$
in dependence on the displacement $R_i$ of the bunch axes. The full and
the dotted curves
correspond to a displacement in the vertical and horizontal direction,
respectively.
\label{cbs3} }
\end{figure}

The effect does not depend on the photon energy and it can be
explained as follows. At $R_{y}=0$ a
considerable portion of the  electrons moves in
the region of small impact parameters where the electric  and
magnetic fields of the positron bunch are small. For $R_{y}$
  in the range of
 $\sigma _{y}^2 \ll  R_{y}^2 \ll \sigma _{x}^2$  the electrons
fly through a stronger  electromagnetic field  of
the  positron  bunch, therefore, the number  of the
emitted photons increases. At large $R_y$ (for $\sigma _x^2 \ll
R_y^2 \ll l^2$, not shown in the Fig.), the fields of the positron bunch
are $|{\bf
E}| \approx |{\bf B}| \propto\; |1/R_y|$ which leads to $dN_\gamma
\propto \; 1/R_y^2$. In that region the number of emitted photons
decreases, but very slowly  compared with the luminosity.

Shifting  the electron bunch axis   in the horizontal direction,
 the positron bunch fields immediately become weaker
and the  photon rate
decreases (see the dotted curve in Fig.~3). However, this decrease is
not so strong as that of
the luminosity. For example, $dN_\gamma (R_x) /dN_\gamma(0) = 0.091$
at $R_x = 5 \,\sigma _{x}$, but the luminosity drops by   three
orders of magnitude.

Such an unusual dependence of the CBS photon rate on
${\bf R}$ can be used for a fast control over
impact parameters between beams and over transverse
beam sizes.  For the case
of long round bunches, such an experiment has already been
successfully performed at the SLC collider \cite{Bon}.

\subsection{Azimuthal asymmetry and polarization}

Taking into account the angular distribution of the final
photons in the Compton cross section (\ref{268}), we obtain instead of
Eq. (\ref{29}) the following distribution of CBS photons
\begin{eqnarray}
dN_\gamma = {3\over 2} N_0 {dE_\gamma \over E_\gamma}
{dz \over (1+z)^4} \  {d\varphi \over 2\pi} \
[1+z^2 - 2z (\xi_3 \cos{2\varphi} \nonumber\\
- \xi_1 \sin{2\varphi})]
\mbox{exp} \left[ - \left( (1+z) {E_\gamma \over E_c}\right)^2 \right] \ .
\label{173}
\end{eqnarray}

If the impact parameter between beams ${\bf R}$ is non--zero, an
azimuthal asymmetry of the CBS photons appears, which   can
be used for an additional control over the beams.
For increasing $R_{x}$  ($R_y$) the electron bunch is
shifted
into the region where the electric field of the positron bunch is
directed almost  horizontally (vertically).  As a result, the
equivalent photons  become
linearly polarized in the direction of the field. In general, the average
degree of the linear polarization is defined by
\begin{equation}
P_l = \sqrt{\xi_1^2 + \xi_3^2}\ .
\label{174}
\end{equation}
For identical beams and horizontal or vertical displacements one
gets from (\ref{267},\ref{23}) that $\xi_1 = 0$.
In Fig.~4, the average longitudinal polarization
for $DA\Phi NE$ is presented.
\begin{figure}[htb]
\epsfig{file=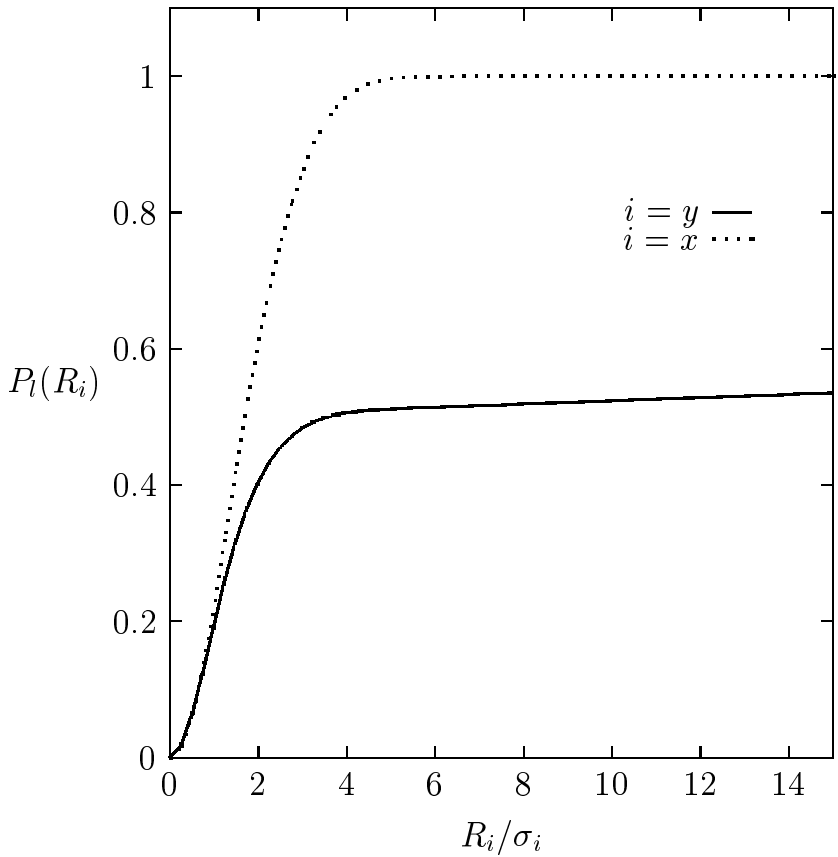,width=12cm}
\caption{Average longitudinal polarization $P_l$ of equivalent photons
at   $DA\Phi NE$
vs. the displacement $R_i$ of the bunch axes. The full and
the dotted curves correspond to vertical and horizontal shifts.
\label{cbs4} }
\end{figure}

Let us define the azimuthal asymmetry of the emitted photons
by the relation
\begin{equation}
A= {dN_{\gamma}(\varphi=0)-dN_{\gamma}(\varphi=\pi/2)\over
dN_{\gamma}(\varphi=0)+dN_{\gamma}(\varphi=\pi/2)}
\label{175}
\end{equation}
where the azimuthal angle $\varphi $ is measured with respect to
the horizontal plane. It is not difficult to obtain that this
quantity does not depend on the photon energy and it is equal to
\begin{equation}
A=\mp {2(\gamma_e \theta)^2 \over 1+(\gamma_e \theta)^4} \;|\xi_3|
\label{176}
\end{equation}
for the horizontal (vertical) displacement of the beams (here
$\theta$ is the polar angle of the emitted photon). From
Fig. 4 one can see that with increasing $R_{x}$ ($R_y$)
the fraction of photons emitted in the vertical (horizontal)
direction becomes greater
than the fraction of photons emitted in the horizontal (vertical)
direction.

For any displacements the equivalent photons are linearly polarized
with the degree
 $P_l$,   the CBS photons are also
linearly polarized in the same direction. Denoting by $P_l^{CBS}$
the average degree of CBS photon polarization, the ratio
$P^{CBS}_l / P_l$ varies in the interval from 0.5 to 1 when
$E_\gamma$ increases \cite{Serbo} (see the Table~\ref{tab1}).

\vspace{0.3cm}
\begin{table}
\caption{\label{tab1} Average degree of the polarization of CBS photons for
different energies.}
\renewcommand{\arraystretch}{1.5}
\begin{center}
 \begin{tabular}{|c|c|c|c|c|}\hline
 $E_\gamma /E_c$ & 0 & 0.2 & 0.4 & 0.6
 \\ \hline
$P^{CBS}_l / P_l$
& 0.5 & 0.7 & 0.81 & 0.86  \\
\hline
 $E_\gamma /E_c$ & 0.8 & 1 & 1.5 & 2
 \\ \hline
$P^{CBS}_l / P_l$
& 0.89 & 0.94 & 0.96 & 0.97  \\
\hline
\end{tabular}
\end{center}
\end{table}

\vspace{1cm}

\noindent
{\bf Acknowledgements}\\
V.G.~Serbo acknowledges   support of the S\"achsisches Staatsministerium
f\"ur Wissenschaft und Kunst, of the
Naturwissenschaftlich--Theoretisches
Zentrum of the Leipzig University and of the Russian
 Fond of Fundamental Research.
R.~Engel is supported by the Deutsche
Forschungsgemeinschaft under grant Schi 422/1-2.

\end{document}